\documentclass[%
 reprint,
 amsmath,amssymb,amsthm,
 nofootinbib,
 aps,
]{revtex4-1}

\usepackage{dcolumn}
\usepackage{bm}
\usepackage[mathlines]{lineno}
\usepackage{graphicx}
\usepackage{csquotes}
\usepackage[final]{changes}

\newtheorem{Theorem}{Theorem}
\newtheorem{Corollary}{Corollary}[Theorem]

\begin{document}
\preprint{APS/123-QED}

\title{Law of Total Probability in Quantum Theory and Its Application in Wigner's Friend Scenario}
\author{Jianhao M. Yang}
\email{jianhao.yang@alumni.utoronto.ca}
\affiliation{Qualcomm, San Diego, CA 92121, USA}

\date{\today}

\begin{abstract}
It is well-known that the law of total probability does not hold in general in quantum theory. However, the recent arguments on some of the fundamental assumptions in quantum theory based on the extended Wigner's Friend scenario show a need to clarify how the law of total probability should be formulated in quantum theory and under what conditions it still holds. In this work, the definition of conditional probability in quantum theory is extended to POVM measurements. Rule to assign two-time conditional probability is proposed for incompatible POVM operators, which leads to a more general and precise formulation of the law of total probability. Sufficient conditions under which the law of total probability holds are identified. Applying the theory developed here to analyze several quantum no-go theorems related to the extended Wigner's friend scenario reveals logical loopholes in these no-go theorems. The loopholes exist as a consequence of taking for granted the validity of the law of total probability without verifying the sufficient conditions. Consequently, the contradictions in these no-go theorems \replaced{only reconfirm the invalidity of the law of total probability in quantum theory, rather than invalidating}{just reconfirm the invalidity of the law of total probability in quantum theorem, rather than confirm} the physical statements that the no-go theorems attempt to refute.
\begin{description}
\item[Keywords] The Law of Total Probability, Extended Wigner's Friend Scenario, PVOM Measurement
\end{description}
\end{abstract}
\pacs{03.65.Ta, 03.65.-w}
\maketitle

\section{Introduction}
\label{intro}
In his seminal paper on the path integral formulation of quantum mechanics~\cite{Feynman}, Feynman started the introduction of his new theory by pointing out that the law of total probability in classical probability theory must be replaced by a new form of rule. Specifically, in a slightly different notation, the classical law of probability, $p(c|a)=\sum_bp(c|b)p(b|a)$ where $p(y|x)$ is the probability of obtaining measurement result $y$ given measurement result $x$, is no longer true in quantum theory, and must be replaced by $\varphi(c|a)=\sum_b\varphi(c|b)\varphi(b|a)$ where $\varphi$ is a complex number called probability amplitude and related to the classical probability by the Born's rule $p(y|x)=|\varphi(y|x)|^2$. From this key idea, Feynman continued to expand the theory that leads to the path integral formulation of quantum mechanics. He also discussed when the new rule of summation over probability amplitude can fall back to the classical law of probability. This is when one ``attempts to perform" intermediate measurements that obtain results of all $b$. In modern terms, what Feynman means ``attempting to perform measurement" here can be understood as the decoherence phenomenon~\cite{Zurek03}. 

The above example shows that it has been long known that the law of total probability cannot be taken for granted in quantum theory. Indeed, many other classical probability rules are only upheld in specific conditions. For instance, a joint probability can be definitely assigned only when the two measurement operators are commutative~\cite{Nielsen, Hayashi, Fine, Malley}; There are many variants of definition of the conditional probability in quantum theory (see a review in \cite{Bobo}). However, a family of no-go theorems recently published\cite{Brukner, Bong, Guerin, Rovelli21} appear to rely on the total law of probability one way or another without considering the sufficient conditions. These no-go theorems are related to the extensively discussed Wigner's Friend experiments. In quantum mechanics, the Wigner's friend~\cite{Wigner,Wigner2} thought experiment has been widely discussed as it tests the validity of many quantum interpretation theories. The significance of such experiment is that Wigner and his friend give two different descriptions of the same physical process happened inside the lab. Deutsch further extended the thought experiment to be applicable to macroscopic system such as the lab system~\cite{Deutsch} itself. Based on that, a more sophisticate extended Wigner's friend experiment is put forwarded by Brukner~\cite{Brukner0, Brukner}. Such experiment setup involves two remotely separated labs. Each lab contains half of an entangled pair of spins and a local observer. Outside each lab there is a super-observer who can choose to perform different types of measurements on the lab as a whole. The intention of such experiment setup is to prove, through a no-go theorem, that measured facts are observer-dependent in quantum theory. Subsequent experiment~\cite{Proietti19} has been carried out to confirm the inequality developed in \cite{Brukner}. A stronger version of no-go theorem is further proposed for reaching the similar conclusion~\cite{Bong}. The statement that measured facts are observer-dependent is considered of importance for the quantum foundation and deserved rigorous theoretical proving and experimental testing. However, proving the no-go theorems by taking the law of total probability for granted casts doubt on the theoretical rigorousness.

The fact that there is still ambiguity to use the total law of probability in quantum theory though it has been long recognized it is not upheld in quantum mechanics shows the needs to provide a rigorous formulation of the law of total probability in quantum theory, and to clarify under what conditions it holds true. And this is indeed the motivation of the present work. The formulation of the law of total probability depends on a clear definition of conditional probability in quantum theory. There is already extensive research literature on how the conditional probability is defined~\cite{Bobo, Bobo2, Luders, Cassinelli, Bub, PW, Dolby, Lloyd, Hohn, Baumann}. However, these formulations are either based on projection measurements, or only considering simultaneous measurements with commutative operators. In this work we will extend a two-time conditional probability formulation from projection measurement to more generic POVM measurements. Generalization for POVM measurement is needed because some of the no-go theorems choose POVM operators in their proofs. We then give several sufficient conditions for the law of total probability to become true. The theory is applied to analyze several no-go theorems related to the extended Wigner's friend scenario. Logical loopholes are shown in these no-go theorems because their proofs rely on the law of total probability one way or another, but the conditions to validate the law are not met. Thus, these no-go theorems do not really prove the results they expect, such as ``measured facts are observer-dependent". Instead, they just indirectly confirm that the law of total probability doesn't not hold in quantum theory.

It is worth to mention that other concerns on these no-go theorems have already been pointed out~\cite{Zukowski, Relano}. In particular, only when a measurement is completed should a probability distribution be assigned. Assigning probability distribution for pre-measurement without result leads to contradiction~\cite{Zukowski}. The analysis in this work will go one step further by showing that even assigning probability distribution for completed measurement, there is still logical loopholes in the no-go theorem. This is because the law of total probability that the proofs rely on does not hold true with the specific measurement operators and initial quantum state being chosen. Lastly, it is important to emphasize that we do not take a stand on the assertions of the no-go theorems themselves. For instance, it could be still a valid statement that “measured facts are observer-dependent”. What we show here is that there are logical loopholes in the proof of the no-go theorems. 

In summary, this paper extends the formulation of conditional probability to generic POVM measurements, and clarifies the conditions under which the law of total probability can be valid in quantum theory. Applying the theory developed in this work to the extended Wigner Friend scenario reveals logical loopholes in several no-go theorems that take for granted on the validity of the law of total probability. The contradictions in these no-go theorems only reconfirm the invalidity of the law of total probability in quantum theorem, rather than \replaced{invalidating the physical statements that the no-go theorems are intended to refute, such as “measured facts are independent of the observer".}{confirm the desired statements such as ``measured facts are observer dependent".} We hope the results presented here can inspire further researches to find more convincing proof and experimental testing. This is important because the implications of the extended Wigner’s friend scenario are conceptually fundamental in quantum theory.

\section{The Law of Total Probability in Quantum Theory}
\label{probTheory}
First, we briefly review the classical probability theory. Suppose there are two random variables X and Y. Without loss of generality, we assume X and Y are discrete random variables. Measuring $X$ (or $Y$) will obtaining one of values in $\{a_i: i=1,2,3...\}$ (or in $\{b_j: j=1,2,3...\}$) which is finite or countable infinite. Denote the joint probability of measuring X with result $X=a_i$ and measuring Y with result $Y=b_j$ as $p(a_i,b_j)$, and the conditional probability of obtaining $X=a_i$ given that $Y=b_j$ as $p(a_i|b_j)$. They are related by the following axioms:
\begin{align}
\label{axiom1}
    p(a_i,b_j) &= p(b_j|a_i)p(a_i)\\
\label{axiom2}
    p(b_j, a_i) &= p(a_i|b_j)p(b_j) \\
\label{axiom3}
    p(a_i,b_j) &= p(b_j, a_i),
\end{align}
where $p(a_i)$ is the marginal probability of measuring X with result $X=a_i$, and similarly for $p(b_j)$. Axiom (\ref{axiom3}) ensures the joint probability is defined uniquely regardless it is defined by (\ref{axiom1}) or (\ref{axiom2}). We explicitly call out (\ref{axiom3}) since it is not always true in the quantum theory. 

The law of total probability can be derived\footnote{Axioms (\ref{axiom1}) to (\ref{axiom3}) give $p(b_j|a_i)p(a_i)=p(a_i|b_j)p(b_j)$, which is the Bayes' law. Summing over $i$ in both sides and using identity $\sum_i p(a_i|b_j)p(b_j)=p(b_j)$, one obtains (\ref{totalLaw2}).} from axioms (\ref{axiom1}) - (\ref{axiom3}) and expressed as following,
\begin{equation}
    \label{totalLaw2}
    p(b_j) = \sum_i p(b_j|a_i)p(a_i).   
\end{equation}
What we want to investigate here is how the equivalent version of (\ref{totalLaw2}) in quantum theory can be formulated.

To start with, we need to examine how the conditional probability is constructed in quantum theory. The subtlety of constructing the conditional probability in quantum theory has been investigated long ago. G. Bobo gives an extensive review and discussion~\cite{Bobo}. The generally accepted formulation of conditional probability in quantum theory is provided by L\"{u}ders rule~\cite{Luders} where the measurements are associated with projection operators. L\"{u}ders rule is based on Gleason's theorem which mathematically justifies Born's rule. Here we wish to follow the similar approach to generalize the formulation for conditional probability when the measurements are associated with POVM operators. 

Mathematical proofs for generalizing Gleason's theorem to POVM measurements are given by \cite{Busch, Caves}, which is our starting point. Suppose a quantum system $S$ is prepared such that its state is described by density operator $\rho$. $S$ could be a composite system, which we will discuss later. Let $A=\{A_i\}$ be a POVM for $S$. The probability of measurement with element $A_i$ resulting in value $a_i$ is~\cite{Busch, Caves}
\begin{equation}
    \label{POVM1}
    p(a_i|\rho)=Tr(\rho A_i),
\end{equation}
and the post-measurement density operator $\rho_i$ are given by \cite{Hayashi}
\begin{equation}
\label{POVM2}
    \rho_i=\frac{\sqrt{A_i}\rho\sqrt{A_i}}{p(a_i|\rho)}.
\end{equation}
Let $B=\{B_j\}$ be another POVM for $S$. Given the post measurement state $\rho_i$, the probability of measurement with element $B_j$ resulting in value $b_j$ is, by applying recursively (\ref{POVM1}), $p(b_j|\rho_i)=Tr(\rho_i B_j)$. Substituting the expression for $\rho_i$ in (\ref{POVM2}), we obtain the conditional probability
\begin{equation}
    \label{cond1}
    p(b_j|a_i,\rho) = \frac{Tr(\sqrt{A_i}\rho\sqrt{A_i} B_j)}{Tr(\rho A_i)}.
\end{equation}
There is an underlying assumption in this definition that the probability is assigned only after the measurements are completed. In particular, the first POVM measurement $A_i$ must be completed in order to be qualified as a condition. We strictly follow this assumption as opposite to assigning a probability with only ``pre-measurement". Pre-measurement refers to only the unitary process that entangles the measured system and measuring apparatus~\cite{Neumann}, but without the projection process to single out a particular outcome.

Given the same initial state $\rho$, if we swap the order of measurements such that $B_j$ goes first, following $A_i$, we obtain a conditional probability
\begin{equation}
    \label{cond2}
    p(a_i|b_j,\rho) = \frac{Tr(\sqrt{B_j}\rho\sqrt{B_j} A_i)}{Tr(\rho B_j)}.
\end{equation}
Note $p(a_i|\rho)=Tr(\rho A_i)$ and $p(b_i|\rho)=Tr(\rho B_j)$, Eqs.(\ref{cond1}) and (\ref{cond2}) can be rewritten as
\begin{align}
    \label{def1}
     p(b_j|a_i,\rho)p(a_i|\rho) &= Tr(\sqrt{A_i}\rho\sqrt{A_i} B_j)\\
    \label{def2}
    p(a_i|b_j,\rho)p(b_j|\rho) &= Tr(\sqrt{B_j}\rho\sqrt{B_j} A_i).
\end{align}
Eqs.(\ref{def1}) and (\ref{def2}) are not necessarily equal, which shows that the quantum version of Bayes' theorem,
\begin{equation}
    \label{Bayes}
    p(a_i|b_j,\rho)p(b_j|\rho)=p(b_j|a_i,\rho)p(a_i|\rho)
\end{equation}
does not hold in general in quantum theory. This posts a difficulty to define a joint probability as either $p(a_i, b_j)=p(a_i|b_j,\rho)p(b_j|\rho)$ or $p(a_i, b_j)=p(b_j|a_i,\rho)p(a_i|\rho)$, because it depends on the order of measurement events. 
Another consequence is that the laws of total probability, i.e., the quantum version of (\ref{totalLaw2})
\begin{equation}
    \label{totalLaw3}
    \sum_ip(b_j|a_i,\rho)p(a_i|\rho) = p(b_j|\rho)
\end{equation}
does not hold either in general. This is because from (\ref{def1}), $\sum_ip(b_j|a_i,\rho)p(a_i|\rho)=\sum_iTr(\sqrt{A_i}\rho\sqrt{A_i} B_j)$, while $p(b_j|\rho)=Tr(\rho B_j)$, and there are not equal in general\footnote{Note that on the other hand, given (\ref{cond1}) and the completeness of POVM elements, $\sum_iA_i=I$ where $I$ is the identity operator, it is straightforward to verify that $\sum_jp(b_j|a_i,\rho)p(a_i|\rho) = p(a_i|\rho)$.}. We are interested in finding the conditions under which (\ref{totalLaw3}) becomes true.

It is well-known that when $[A_i, B_j]=0$, i.e., $A_i$ and $B_j$ commute, from (\ref{cond1}) and (\ref{cond2}), one gets $p(b_j|a_i,\rho)p(a_i|\rho)=p(a_i|b_j,\rho)p(b_j|\rho)=Tr(\rho A_iB_j)$. Consequently, the law of total probability (\ref{totalLaw3}) becomes true and a joint probability can be well-defined. However, the situation becomes much complicated when $[A_i, B_j]\ne 0$.

Strictly speaking, due to the uncertainty principle, when $A_i$ and $B_j$ are non-commutative, the two measurements cannot be performed to obtain definite outcomes at the same time. The conditional probability defined in (\ref{cond1}) or (\ref{cond2}) needs to be extended to a two-time formulation of conditional probability in order to be applicable when $[A_i, B_j]\ne 0$. There are extensive research literature on how to construct two-time conditional probability in quantum theory~\cite{Bobo, Bobo2, Luders, Cassinelli, Bub, PW, Dolby, Lloyd, Hohn, Baumann}. One noticeable approach is based on the Page-Wootters' timeless formulation~\cite{PW, Dolby, Lloyd, Hohn, Baumann}. In this work, we will continue to be based on the generalized Gleason theorem for POVM~\cite{Busch, Caves} to derive the two-time conditional probability, and leave the discussion of the Page-Wootters mechanism in Section IV.

For conceptual clarity, we start the analysis by considering that there is finite non-zero duration for each measurement. After we construct the conditional probability formulation, for practical purpose of calculation, we can approximate the measurement duration to zero. Suppose the first measurement starts at $t_a^-$ and completes at $t_a^+$. Here $t_a^+-t_a^-$ covers the time duration for both the pre-measurement unitary phase that entangles the measured system and the measuring apparatus, and the projection phase. The measurement process\footnote{Theorem 5.2 of \cite{Hayashi} gives a detailed account on how this POVM measurement is physically realized through indirect measurement.} is represented by a POVM element $A_i$ associated with outcome $a_i$. Similarly, the second measurement starts at $t_b^-$ and completes at $t_b^+$. Between $t_a^+$ and $t_b^-$ is a free time evolution for the measured system $S$, described by operator $U(\Delta t)=e^{-iH\Delta t/\hbar}$ where $\Delta t=(t_b^--t_a^+)$. Since it is only meaningful to assign a probability distribution after a measurement is completed, the two-time conditional probability we want to construct is ``given the measurement outcome of $a_i$ at $t_1$ where $t_a^+ < t_1 < t_b^-$, what is the probability of measurement outcome $b$ at $t_2 > t_b^+$. Mathematically, this two-time conditional probability can be written as $p(b_j\text{ at } t_2|a_i\text{ at } t_1, \rho_0)$, where $\rho_0\equiv\rho(t_a^-)$ is the initial density operator of $S$ when the first measurement starts. After the first measurement with POVM element $A_i$, the post-measurement state is $\rho_i(t_a^+)=\sqrt{A_i}\rho_0\sqrt{A_i}/Tr(\rho_0 A_i)$. The quantum system $S$ then time evolves from $t_a^+$ to $t_b^-$ to a new state $\rho_i(t_b^-)=U(\Delta t)\rho_i(t_a^+)U^\dag(\Delta t)$. At $t_b^-$, the second measurement occurs. This is represented by applying POVM element $B_j$ on $\rho_i(t_b^-)$ and obtain outcome $b_j$ at $t_b^+$ with probability $Tr(B_j\rho_i(t_b^-))$. Substituting $\rho_i(t_b^-)$, the two-time conditional probability is
\begin{equation}
\label{twotimes}
    \begin{split}
    &p(b_j\text{ at } t_2|a_i\text{ at } t_1, \rho_0) =Tr(B_j\rho_i(t_b^-)) \\
    &=\frac{Tr(B_jU(\Delta t)\sqrt{A_i}\rho_0\sqrt{A_i}U^\dag (\Delta t))}{Tr(\rho_0 A_i)}.
    \end{split}   
\end{equation}

For practical purpose of calculation, we can assume the measurement duration is very small compared to the free evolution time, i.e., $(t_a^+-t_a^-) \ll \Delta t$ and $(t_b^+ - t_b^-) \ll \Delta t$. Then we can denote  $t_a^-\approx t_a^+$ as $t_a$, $t_b^-\approx t_b^+$ as $t_b$, and $\Delta t = (t_b-t_a)$. 

Suppose the two POVM elements $A_i$ and $B_j$ are projection measurements, $A_i=|\phi_i\rangle\langle \phi_i|$ and $B_j=|\varphi_j\rangle\langle \varphi_j|$, one can verify that the conditional probability defined in (\ref{twotimes}) gives the correct transition probability in standard quantum mechanics:
\begin{equation}
    \label{transProb}
    p(b_j\text{ at } t_2|a_i\text{ at } t_1, \rho_0) = |\langle\phi_i|U(\Delta t)|\varphi_j|^2.
\end{equation}
However, Eq.(\ref{twotimes}) is more generic as it is defined with general POVM operators.
Note that the denominator in (\ref{twotimes}) $Tr(\rho_0 A_i)=p(a_i\text{ at } t_1 |\rho_0)$, (\ref{twotimes}) can be rewritten as
\begin{equation}
    \label{def3}
    \begin{split}
    &p(b_j\text{ at } t_2|a_i\text{ at } t_1, \rho_0)p(a_i\text{ at } t_1 |\rho_0) \\
    &=Tr(B_jU(\Delta t)\sqrt{A_i}\rho_0\sqrt{A_i}U^\dag (\Delta t)).
    \end{split}
\end{equation}
To analyze the two-time version of total law of probability, which can be expressed as 
\begin{equation}
    \label{totalLaw5}
    \begin{split}
    p(b_j\text{ at } t_2|\rho_0) 
    &=\sum_i p(b_j\text{ at } t_2|a_i\text{ at } t_1, \rho_0)p(a_i\text{ at } t_1 |\rho_0),
    \end{split}
\end{equation}
we consider a series of two-time measurements $\{A_i \text{ at } t_a, B \text{ at }t_b, i=1\ldots N\}$ on $N$ copies of measured system $S$ with the same initial state $\rho_0$. Each two-time measurement consists a first measurement from one possible POVM element from the complete set $\{A_i, i=1\ldots N\}$ at time $t_a$, and the same second measurement $B_j$ at time $t_b$. For $t_a < t_1 < t_b < t_2$, from (\ref{def3}) we have
\begin{equation}
    \label{def4}
    \begin{split}
    &\sum_i p(b_j\text{ at } t_2|a_i\text{ at } t_1, \rho_0)p(a_i\text{ at } t_1 |\rho_0) \\
    &=\sum_i Tr(B_jU(\Delta t)\sqrt{A_i}\rho_0\sqrt{A_i}U^\dag(\Delta t)).
    \end{split}
\end{equation}
But by definition, $p(b_j\text{ at } t_2|\rho_0) = Tr(B_jU\rho_0U^\dag)$. We can see (\ref{totalLaw5}) is not true in general. The Theorem next attempts to address the question on under what conditions (\ref{totalLaw5}) is valid. 
\begin{Theorem} 
\label{theorem1}
Let $\rho_0$ be the density operator for a quantum system $S$ before the measurements. Let $A_i$ and $B_j$ be two POVM elements to measure $S$ at time $t_a$ and $t_b$ respectively, and $U(t_b, t_a)$ is the unitary time evolution operator from $t_a$ to $t_b$. Select $t_1$ and $t_2$ such that $t_a < t_1 < t_b < t_2$. The law of total probability (\ref{totalLaw5}) is true if one of the following conditions is met.
\begin{enumerate}
    \item[C1.] $[A_i, U^{\dag}B_jU] = 0$, $\forall \rho_0$, 
    \item[C2.] $\rho_0=\sum_i\lambda_i|\phi_i\rangle\langle \phi_i|$ and $A_i=|\phi_i\rangle\langle \phi_i|$,
    \item[C3.] $\rho_0$ is a pure state, given by $|\Psi\rangle\langle\Psi|$, $A_i$ is a projection operator and $\langle\Psi|[A_i,U^{\dag}B_jU]A_i|\Psi\rangle=0$.
\end{enumerate}
\end{Theorem}
The proof of Theorem \ref{theorem1} is left in Appendix A, but a few comments are in order here. First, Condition $C1$ implies $U^{\dag}B_jUA_i=A_iU^{\dag}B_jU$. The sequence of operations for $U^{\dag}B_jUA_i$ means performing measurement $A_i$ at $t_a$, time evolving the post-measurement state from $t_a$ to $t_b$, performing measurement $B_j$ at $t_b$, and reversing time evolution of the post-measurement state back to $t_a$. The sequence of operations $A_iU^{\dag}B_jU$ means time evolving the state from $t_a$ to $t_b$, performing measurement $B_j$ at $t_b$, then reversing time evolution of the state back to $t_a$, and performing measurement $A_i$ at $t_a$. Condition $C1$ says that if these two sequences of operations are equivalent, then the law of total probability (\ref{totalLaw5}) holds true.

Second, if the post-measurement state $\rho_i(t_a)$ after the first measurement does not change during free time evolution, such as the case of a spin state in free space, we will have $\rho_i(t_b)=\rho_i(t_a)=\sqrt{A_i}\rho_0\sqrt{A_i}/Tr(\rho_0 A_i)$. Then, (\ref{twotimes}) can be written as
\begin{equation}
\label{twotimes2}
     p(b_j\text{ at } t_2|a_i\text{ at } t_1, \rho) = \frac{ Tr(\rho_0\sqrt{A_i}B_j\sqrt{A_i})}{Tr(\rho_0 A_i)}.
\end{equation}
Eq.(\ref{twotimes2}) appears the same as (\ref{cond1}), but the precisely meaning is different in that the two measurements $A_i$ and $B_j$ in (\ref{twotimes2}) are taken place at two different times. With such special post-measurement quantum state, the sufficient conditions in Theorem \ref{theorem1} become
\begin{enumerate}
    \item[$\textit{C1}^{\prime}$.] $[A_i, B_j] = 0$, $\forall \rho_0$, 
    \item[$\textit{C2}^{\prime}$.] $\rho_0=\sum_i\lambda_i|\phi_i\rangle\langle \phi_i|$ \textit{and} $A_i=|\phi_i\rangle\langle \phi_i|$,
    \item[$\textit{C3}^{\prime}$.] $\rho_0=|\Psi\rangle\langle\Psi|$, $A_i$ \textit{is a projection operator and} $\langle\Psi|[A_i,B_j]A_i|\Psi\rangle=0$.
\end{enumerate}

We close this section with two comments. First, when two measurement operations are not commutative, the conditional probability needs to be defined in the two-time formulation. Second, We can give an intuitive explanation of why (\ref{totalLaw5}) does not hold in general in quantum theory. As shown in (\ref{def4}), the right hand side of (\ref{totalLaw5}) refer to the summation of traces of multiplication of operators from a series of experiments where two measurements are carried out in a sequence. In the case of special post-measurement state where (\ref{twotimes2}) holds, this is $\sum_i Tr(B_j\sqrt{A_i}\rho_0\sqrt{A_i})$. The measurement of $A_i$ changes the initial quantum state such that it affects the probability of outcome for a subsequent measurement $B_j$. However, the term on the left hand side of (\ref{totalLaw5}) refers to the probability of an experiment where only measurement $B_j$ is carried out with the same initial quantum state. There is no reason to assume both sides are equal. Eq.(\ref{totalLaw5}) holds only in special conditions such as those specified in Theorem \ref{theorem1}. 

The conclusion here is that one should not take for granted that the law of total probability holds true in general. Instead, the sufficient conditions, such as those provided in Theorem \ref{theorem1}, need to be clearly called out. Failing to do so may leave loophole in logical deduction.

\begin{figure*}
\begin{center}
\includegraphics[scale=2.75]{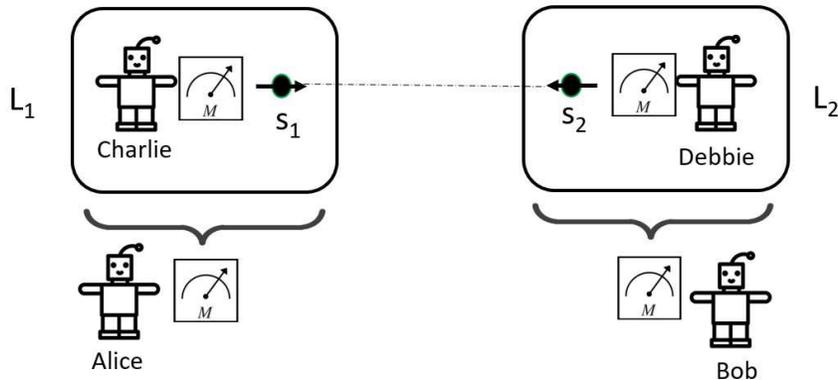}
\caption{Sketch of the extended Wigner's Friend scenario described in~\cite{Brukner}. Laboratory $L_1$ consists spin $s_1$ and Charlie, while Laboratory $L_2$ consists spin $s_2$ and Debbie. The two laboratories are remotely separated. The dot line between $s_1$ and $s_2$ symbolizes they are entangled. Alice can measure $L_1$ as a whole, and Bob can measure $L_2$.}
\label{fig:1}       
\end{center}
\end{figure*}
\section{Application to Composite Systems}
\label{sub:composite}
In this subsection, we will apply the conditional probability definition to composite quantum systems, and re-examine Theorem \ref{theorem1} when measuring composite system. Suppose the measured system $S$ consists two subsystems $S_1$ and $S_2$ that are space-like separated. Define $A_i = P_i\otimes I_2$ where $P_i= |\phi_i\rangle\langle \phi_i|$ is a local POVM element on subsystem $S_1$ and $I_2$ is an identity operator on subsystem $S_2$. Similarly, define $B_j = I_1\otimes Q_j$ where $Q_j$ is a local POVM elements on subsystem $S_2$. By the principle of locality, a local measurement on a subsystem should not cause any impact on the other remote subsystem. Therefore, $[A_i, B_j] = 0$. For measurement outcomes of such two local measurements, Eqs.(\ref{cond1}) and (\ref{cond2}) are correct formulations for conditional probability, the joint probability is well-defined. Consequently, Eqs.(\ref{Bayes}) and (\ref{totalLaw3}) hold true. There is no need to use the two-time formulation of conditional probability. This is the case for typical Bell tests and has been used in deriving the Bell-CHSH inequalities\footnote{However, in the derivation of Bell-CHSH inequalities, identity (\ref{axiom1}) is further expressed as 
\begin{equation}
    \label{OI}
    p(a_i,b_j|\lambda) = p(a_i|b_j,\lambda)p(b_j|\lambda) = p(a_i|\lambda)p(b_j|\lambda),
\end{equation}
where $\lambda$ is a hidden variable. This is known as the \textit{Outcome Independence} assumption~\cite{Hall, Hall2}.}.

However, suppose $B_j = Q_j\otimes I_2$ where $Q_j$ is another local POVM elements on subsystem $S_1$ and $[P_i, Q_j] \ne 0$. In this case, Eq.(\ref{cond1}) is incorrect for conditional probability. The two-time conditional probability formulation is needed and can be calculated as
\begin{equation}
\label{twotimes10}
    \begin{split}
    &p(b_j\text{ at } t_2|a_i\text{ at } t_1, \rho_0) \\
    &=\frac{Tr((Q_j\otimes I_2) U(\Delta t)\sqrt{P_i\otimes I_2}\rho_0\sqrt{P_i\otimes I_2}U^\dag (\Delta t))}{Tr(\rho_0 P_i\otimes I_2)},
    \end{split}   
\end{equation}
where $U(\Delta t)=U_{S_1}(\Delta t)\otimes U_{S_2}(\Delta t)$.

Next we wish to apply the two-time conditional probability to the extended Wigner's friend (EWF) scenario introduced in~\cite{Brukner}. As shown in Figure \ref{fig:1}, the EWF scenario consists two space-like separated laboratories $L_1$ and $L_2$. Each laboratory contains half of an entangled pair of systems $s_1$ and $s_2$. $L_1$ also contains a friend Charlie who can perform measurements on $s_1$. Outside $L_1$ there is a super-observer Alice who can perform different types of measurements on $L_1$ as a whole.  Similarly, there is a friend Debbie in $L_2$, and a super-observer Bob outside $L_2$. Here, we need to consider four POVM measurements, represented by POVM elements $A, B, C, D$. Here operators $A$ and $C$ act on Hilbert space $\mathcal{H}_{L_1}$, and $B$ and $D$ act on Hilbert space $\mathcal{H}_{L_2}$. We drop the subscripts of the operators and $\rho_0$ for simplifying notations. In a typical EWF experiment, the chosen operators are not all commutative with one another. Specifically, $[A, C]\ne 0$ and $[B, D]\ne 0$, while $[C\otimes I_{L_2}, I_{L_1}\otimes D]=0$ and $[A\otimes I_{L_2}, I_{L_1}\otimes B]=0$. The two-time probability formulation to compute the conditional probability is needed because the measurements $C$ and $D$ are taken place before the measurements of $A$ and $B$. Since $[C\otimes I_{L_2}, I_{L_1}\otimes D]=0$ and $[A\otimes I_{L_2}, I_{L_1}\otimes B]=0$, we can assume measurement $C$ and $D$ are taken place at the same time, $t_a$, as $C\otimes D$ while measurements of $A$ and $B$ at a same later time $t_b$ as $A\otimes B$. Without loss of clarity, we drop the symbol $\otimes$ hereafter. Then the conditional probability for $t_a < t_1 < t_b < t_2$ is given by
\begin{equation}
    \label{4ops}
    \begin{split}
     &p(ab \text{ at } t_2|cd \text{ at }t_1, \rho, xy) \\ &=\frac{Tr(\rho\sqrt{C}\sqrt{D}U^{\dag}ABU\sqrt{C}\sqrt{D})}{Tr(\rho CD)},
     \end{split}
\end{equation}
where $U=U_{L_1}\otimes U_{L_2}$ is the time evolution operator from $t_a$ to $t_b$. The law of total probability we are interested is
\begin{equation}
    \label{totalLaw11}
    \begin{split}
        &p(ab  \text{ at }t_2|\rho, xy) =\\
        &\sum_{cd}p(ab \text{ at } t_2|cd \text{ at }t_1, \rho, xy)p(cd  \text{ at }t_1|\rho, xy)\\
    \end{split}
\end{equation}
From (\ref{4ops}), the R.H.S. of (\ref{totalLaw11}) becomes
\begin{equation}
    \label{RHS7}
    \begin{split}
     &\sum_{cd}p(ab \text{ at } t_2|cd \text{ at }t_1, \rho, xy)p(cd  \text{ at }t_1|\rho, xy)\\ &=\sum_{CD}Tr(\rho\sqrt{C}\sqrt{D}U^{\dag}ABU\sqrt{C}\sqrt{D}).
     \end{split}
\end{equation}
The summation is over POVM element sets for $\{C\}$ and $\{D\}$. Since $[A\otimes I_{L_2}, I_{L_1}\otimes B]=0$, the L.H.S. of (\ref{totalLaw11}) is $p(ab  \text{ at }t_2|\rho, xy)=Tr(\rho U^{\dag}ABU)$. Both sides are not equal in general. 

In the case that the post-measurement state after the first measurement is unchanged during free time evolution\footnote{This is indeed the assumption in the no-go theorems we will analyze in next section.}, Eq.(\ref{4ops}) becomes
\begin{equation}
    \label{4ops2}
     p(ab \text{ at } t_2|cd \text{ at }t_1, \rho, xy) =\frac{Tr(\rho\sqrt{C}\sqrt{D}AB\sqrt{C}\sqrt{D})}{Tr(\rho CD)}.
\end{equation}
Eq.(\ref{RHS7}) is simplified to
\begin{equation}
    \label{RHS8}
    \begin{split}
     &\sum_{cd}p(ab \text{ at } t_2|cd \text{ at }t_1, \rho, xy)p(cd  \text{ at }t_1|\rho, xy)\\ &=\sum_{CD}Tr(\rho\sqrt{C}\sqrt{D}AB\sqrt{C}\sqrt{D}),
     \end{split}
\end{equation}
and $p(ab  \text{ at }t_2|\rho, xy)=Tr(\rho AB)$. In this case, one can derive the following corollary based on Theorem \ref{theorem1}.
\begin{Corollary} 
\label{corollary2}
In the Extended Wigner's Friend scenario setup, suppose the post-measurement state is unchanged during free time evolution from $t_a$ to $t_b$. Select $t_1$ and $t_2$ such that $t_a < t_1 < t_b < t_2$. The law of total probability (\ref{totalLaw11}) is true if one of the following conditions is met.
\begin{enumerate}
    \item[C4.] $[A, C]=0$ and $[B, D]=0$, $\forall \rho_0$, 
    \item[C5.] $\rho_0=|\Psi\rangle\langle\Psi|$, $C$ and $D$ are projection operators, and
    \begin{equation}
    \label{C34}
    \langle\Psi|[CD, AB]CD|\Psi\rangle=0.
    \end{equation}
\end{enumerate}
\end{Corollary}
Condition $C4$ is quite obvious. Proof of condition (\ref{C34}) is given in Appendix B.

\section{Logical Loopholes in No-go Theorems Related to the Wigner's Friend Scenario}
A no-go theorem usually starts from the conventional probability theory, which is widely regarded as the true representation of logical deduction, and assumes certain additional plausible physical premises: realism, locality, no superdeterminism, observer independence, etc. One then shows that such a model leads to prediction which is contradicted by quantum mechanics. Hence, one concludes that, at least one of the assumptions, or the rules of conventional probability must be violated by quantum mechanics. Let's denote the physical assumption that a no-go theorem tries to prove to be violated by quantum theory as $\mathcal{W}$. For instance, $\mathcal{W}$ could be ``measured facts are observer independent". The no-go theorem may be constructed independent of the underlying physical theory. But if the logical deduction in the proof of theorem utilizes the law of total probability in one of the forms of (\ref{totalLaw3}), (\ref{totalLaw5}), or (\ref{totalLaw11}) without calling out the appropriate sufficient condition $C$, then we know the resulting statement (could be in a form of inequality) will not hold in quantum theory. This leaves a loophole in the logical deduction. Because the contradiction shown in the no-go theorem could be just due to the fact that $C$ is not met in the experiment setup, instead of the intended conclusion that $\mathcal{W}$ is violated by quantum theory. Thus, the no-go theorem does not reach the conclusion as desired. We will examine several such no-go theorems in this section\footnote{We do not include the no-go theorem~\cite{Renner} widely discussed in the literature since its proof does not invoke the law of total probability.}.

\subsection{A strong no-go theorem on the Wigner’s friend paradox}
\label{Sub:Bong}
Bong \textit{et al.} introduce a no-go theorem that assuming quantum mechanics is applicable to the scale of an observer, then one of the three assumptions, `Locality', `No-superdeterminism', or `Absoluteness of Observed Events (AOE)' must be false~\cite{Bong}. Here AOE means that ``every observed event exists absolutely, not relative to anything or anyone". The no-go theorem is supposed to be independent of underlying physical theory and is proved in the context of extended Wigner's friend (EWF) scenario~\cite{Brukner}, as shown in Figure \ref{fig:1}. The measurement results from the friend in the lab can be correlated with the super-observer's subsequent measurement results. Suppose the measurement outcomes from Alice, Bob, Charlie, and Debbie are $a, b, c, d$ respectively. Alice can have three different measurement settings, labeled by parameter $x \in \{1,2,3\}$. When $x=1$, Alice opens the $L_1$ and asks Charlie's measurement outcome, while when $x=2,3$, Alice performs a different measurement on $L_1$. Similar measurement settings for Bob are labeled as $y\in \{1,2,3\}$.

Among the three assumptions, it is more well accepted that `Locality' and `No-superdeterminism' cannot be violated by any physical theory. The focus is on the assumption of AOE, which is defined mathematically as following~\cite{Bong}. There exists a joint probability distribution $p(abcd|xy)$ such that
\begin{enumerate}
    \item[i] $p(ab|xy)=\sum_{cd}p(abcd|xy) \forall a, b, x, y$
    \item[ii] $p(a|cd, x=1, y)=\delta_{a,c} \forall a,c,d,y$
    \item[iii] $p(b|cd, x, y=1)=\delta_{b,d} \forall b,c,d,x$.
\end{enumerate}
With the three assumptions, ~\cite{Bong} derives a number of inequalities and experimentally confirms that quantum theory violates these inequalities when proper measurement settings $x$ and $y$ and initial quantum state are chosen. Therefore, the AOE assumption should be refuted. 

However, closer examination of the derivation shows that the no-go theorem assumes the law of total probability. The definition of AOE states that $p(ab|xy)=\sum_{cd}p(abcd|xy)$. Then, in Eq.(3) of Ref.~\cite{Bong}, it implicitly assumes $p(abcd|xy) = p(ab|cdxy)p(cd|xy)$. Together, they imply
\begin{equation}
\label{totalLaw7}
    p(ab|xy)=\sum_{cd}p(ab|cdxy)p(cd|xy).
\end{equation}
But as discussed in Section \ref{probTheory}, the law of total probability does not hold true in quantum theory unless certain condition is met.

We can apply Corollary \ref{corollary2} to analyze the validity of (\ref{totalLaw7}). In Appendix C, we show that the operators chosen in Ref.~\cite{Bong} are not all commutative with each other. Specifically, $[A, C]\ne 0$ and $[B, D]\ne 0$, while $[C\otimes I_{L_2}, I_{L_1}\otimes D]=0$ and $[A\otimes I_{L_2}, I_{L_1}\otimes B]=0$. With these choices of operators, the corresponding two-time version of the law of total probability is given by (\ref{totalLaw11}) and (\ref{RHS8}). However, we already see condition $C4$ is not satisfied. The choice of initial quantum state, i.e., Eq.(1) in Ref.~\cite{Bong} and the forms of operator $A, B, C, D$ do not satisfy conditions $C5$ in Corollary \ref{corollary2} either. 

Therefore, in general (\ref{totalLaw7}) does not hold with the conditions specified in \cite{Bong}. Inequalities derived based on (\ref{totalLaw7}) will be violated by quantum theory with the choice of initial quantum state and measurement operators described in~\cite{Bong}. But this raises the question of exactly what the no-go theorem refutes. We agree with the authors that the violation of the inequalities by quantum theory points to the validity of AOE in quantum theory. But the definition of AOE and the derivation of the theorem implicitly assume the validity of the law of total probability. The root cause of the violation of the inequalities is due to the fact that the experimental setup does not satisfy the conditions to make the law of total probability hold true, not because of the AOE statement that ``an observed event is not relative to anything or anyone". One may argue that the AOE statement is equivalent to the invalidity of the law of total probability. But as discussed earlier, the invalidity of  the law of total probability is due to the fact that measurement $C$ (or $D$) alters the initial quantum state that impacts the probability of outcome for measurement $A$ (or $B$) since $[A, C]\ne 0$ (or $[B, D]\ne 0$). There is a logical gap to equate this reason with the statement ``an observed event is not relative to anything or anyone". Therefore, it appears that the violation of inequalities in \cite{Bong} just reconfirms the consequence of non-commutative measurements and therefore the invalidity of the law of total probability in quantum theory, instead of the AOE statement.

\subsection{A no-go theorem for observer-independent facts}
The no-go theorem for observer-independent facts by Brukner~\cite{Brukner} was actually introduced earlier than~\cite{Bong} in a similar effort to prove that measured facts are observer-dependent. The experimental setup is shown in Figure \ref{fig:1}. 
In \added{Ref.}~\cite{Brukner}, there are only two different measurement setups either Alice or Bob will perform, compared to three different measurement setups in \added{Ref.}~\cite{Bong}. Furthermore, the way the no-go theorem in is proved is different. The proof in~\cite{Brukner} leverages the well-known Bell-CHSH inequality thus inherits all the assumptions associated with the Bell-CHSH theorem, while the no-go theorem in ~\cite{Bong} is proved independent of Bell-CHSH theorem. As we will explain next, the proof in~\cite{Brukner} is more subtle, as it carefully chooses a quantum state and a set of measurement operators such that the law of total probability holds true if only considering Alice's (or Bob's) measurements.

The initial wave function is chosen such that after Charlie and Debbie each perform a measurement of the respective half of entangled spin along $z$ axis, from Alice's or Bob's perspective, it becomes (see Eqs.(5)-(9) in ~\cite{Brukner}):
\begin{equation}
    \label{BruknerWF}
    \begin{split}
    |\Psi\rangle&= - \frac{1}{\sqrt{2}}\sin\frac{\theta}{2}(|00\rangle_{L_1}|00\rangle_{L_2} + |11\rangle_{L_1}|11\rangle_{L_2})\\
    &+\frac{1}{\sqrt{2}}\cos\frac{\theta}{2}(|00\rangle_{L_1}|11\rangle_{L_2} - |11\rangle_{L_1}|00\rangle_{L_2}),
    \end{split}
\end{equation}
where $|00\rangle_{L_2}$ represents that $s_1$ is in the spin up state and Charlie's pointer variable associated with the up state, and $|11\rangle_{L_1}$ corresponds to the spin down state for $s_1$ and Charlie's pointer variable. Similar meanings for $|00\rangle_{L_2}$ and $|11\rangle_{L_2}$ for $s_2$ and Debbie's pointer variable. The key point of ~\cite{Brukner} is to assume there exists a joint probability $p(a_1a_2b_1b_2)$, where $a_1, a_2\in \{0, 1\}$ are the measurement results corresponding to Alice's choice of two types of measurement operations and $b_1, b_2\in \{0, 1\}$ are the measurement results of Bob. Alice can choose to either measure $L_1$ with projection operator $A_1$, or with projection operator $A_2$. Here $A_1$ is represented\footnote{In \cite{Brukner} the two types of operations for Alice are defined as $\mathcal{A}_1=|00\rangle\langle 00|_{L_1}-|11\rangle\langle 11|_{L_1}, i\in \{0,1\}$, and $\mathcal{A}_2=|00\rangle\langle 11|_{L_1}-|00\rangle\langle 11|_{L_1}$. Here we use the spectral decomposition theorem to decompose $\mathcal{A}_1$ into projection operators, and represent it by $A_1$. Similar approach for the definition of $A_2$. There is an important difference here compared to the setup in \cite{Bong}. Here $a_1, a_2$ are results for Alice from completed measurement $A_1, A_2$ respectively, whereas in \cite{Bong} $a, c$ are measurement results for Alice and Charlie respectively. The issue of $c$ as result of Charlie's ``pre-measurement" in \cite{Bong} does not exist here in \cite{Brukner}.} by $|\phi_i\rangle\langle\phi_i|_{L_1}$, and $|\phi_0\rangle=|00\rangle$ for $a_1=0$ or $|\phi_1\rangle=|11\rangle$ for $a_1=1$. $A_2$ is chosen to be $A_2=|\chi_0\rangle\langle\chi_0|$ for $a_2=0$ or $A_2=|\chi_1\rangle\langle\chi_1|$ for $a_2=1$, where 
\begin{equation}
    |\chi_i\rangle = \frac{1}{\sqrt{2}}((-1)^i|00\rangle_{L_1} + |11\rangle_{L_1}).
\end{equation}
From these definition of $A_1$ and $A_2$, one can verify that
\begin{equation}
    \label{nonCM}
    [A_1, A_2] = (-1)^{a_1+a_2}(|00\rangle\langle 11|_{L_1} - |11\rangle\langle 00|_{L_1}).
\end{equation}
Similar definitions for operators $B_1$ and $B_2$. The problem in Ref.~\cite{Brukner} is that it assumes the law of marginal probability holds true, for instance, $p(a_2b_2)=\sum_{a_1,b_1=\{0,1\}}p(a_1a_2b_1b_2)$. Ref.~\cite{Brukner} does not provide details on how the joint probability is defined. As discussed earlier, the joint probability cannot be well defined unless the measurement operators are commutative. If we further assume the validity of the classical probability axiom in Eq. (1) and apply it recursively, we have 
\begin{equation}
\label{joint4}
\begin{split}
        p(a_1a_2b_1b_2) &= p(a_2b_1b_2|a_1)p(a_1) \\
        &= p(a_2b_2|a_1b_1)p(b_1|a_1)p(a_1) \\
        &= p(a_2b_2|a_1b_1)p(a_1b_1),
\end{split}
\end{equation}
then the law of marginal probability is equivalent to the law of total probability such that
\begin{equation}
    \label{totalLaw8}
    p(a_2b_2) = \sum_{a_1,b_1=\{0,1\}}p(a_2b_2|a_1b_1,\rho)p(a_1b_1).
\end{equation}
Let's analyze if Eqs. (\ref{joint4}) and (\ref{totalLaw8}) hold true with the chosen operators $\{A_1, B_1, A_2, B_2\}$ and the quantum state in (\ref{BruknerWF}). To do this, we
apply Corollary \ref{corollary2} by replacing operators $\{A, B, C, D\}$ with operators $\{A_2, B_2, A_1, B_1\}$, respectively, and setting $\rho_0=|\Psi\rangle\langle\Psi|$ where $|\Psi\rangle$ is defined in (\ref{BruknerWF}). The conditional probability is similar to (\ref{4ops2}), 
\begin{equation}
    \label{4ops3}
    \begin{split}
     &p(a_2b_2 \text{ at } t_2|a_1b_1 \text{ at }t_1, \rho, xy) \\ &=\frac{Tr(\rho\sqrt{A_1}\sqrt{B_1}A_2B_2\sqrt{A_1}\sqrt{B_1})}{Tr(\rho A_1B_1)},
     \end{split}
\end{equation}
where we drop the subscript of $\rho_0$. The desired law of total probability is
\begin{equation}
    \label{totalLaw12}
    \begin{split}
    &p(a_2b_2 \text{ at } t_2|\rho) = \\
    &\sum_{a_1,b_1}p(a_2b_2 \text{ at } t_2|a_1b_1 \text{ at }t_1, \rho)p(a_1b_1 \text{ at }t_1| \rho).
    \end{split}
\end{equation}
In Appendix D, we show that for the choices of the set of operators $\{A_1, A_2, B_1, B_2\}$ prescribed earlier, $[A_1, A_2]\ne 0$ and $[B_1, B_2]\ne 0$. With the quantum state (\ref{BruknerWF}), Condition $\textit{C3}^{\prime}$ is satisfied such that $p(a_2|\rho)=\sum_{a_1}p(a_2|a_1,\rho)p(a_1|\rho)$ and $p(b_2|\rho)=\sum_{b_1}p(b_2|b_1,\rho)p(b_1|\rho)$. But unfortunately, we also show that  (\ref{totalLaw12}) and the law of marginal probability $p(a_2b_2)=\sum_{a_1,b_1=\{0,1\}}p(a_1a_2b_1b_2)$ are still not valid.

Since the prove of the no-go theorem in \cite{Brukner} depends on the law of marginal probability, and the law of marginal probability does not hold true by the choice of quantum state and measurement operators, there is a logical loophole in the no-go theorem. The violation of the inequality in \cite{Brukner} in quantum theory does not necessarily imply that measured facts are observer dependent. Instead, the violation just reconfirms that the law of marginal probability does not hold for the choice of quantum state and the measurement operators. The logical gap to equate the statement that ``measured facts are observer dependent" to the invalidity of the law of marginal probability in quantum theory is similar to what we discussed in the last paragraph in Section \ref{Sub:Bong}. 

Note that besides depending on the law of marginal probability, the proof of no-go theorem in \cite{Brukner} also inherits the assumptions for the proof of the Bell-CHSH inequality~\cite{Hall, Hall2}, particularly the dependency on the outcome independence assumption (\ref{OI}). The no-go theorem in \cite{Bong}, on the other hand, does not depend on the outcome independence assumption.

\subsection{A no-go theorem for the persistent reality of Wigner’s friend's perception}

In \added{Ref. }\cite{Guerin}, another no-go theorem is introduced to show that in the extended Wigner's friend scenario, Wigner's friend cannot ``treat her perceived measurement outcome as having reality across multiple times" without contradicting one of the following assumptions in quantum mechanics~\cite{Guerin}.
\begin{enumerate}
    \item[P1] Let $f_1$ and $f_2$ be perceived measurement records of the friend at time $t_1$ and $t_2$, respectively. A joint probability distribution $p(f_1, f_2)$ can be assigned that also satisfies the law of total probability $p(f_1)=\sum_{f_2}p(f_1, f_2)$ and $p(f_2)=\sum_{f_1}p(f_1, f_2)$;
    \item[P2] One time probability is assigned according to $p(f_i)=Tr(|f_i\rangle\langle f_i|\rho)$ using unitary quantum theory where no state collapse is considered occurred;
    \item[P3] The joint probability of the friend's perceived outcomes $p(f_1, f_2)$ has a convex linear dependence on the initial state $\rho$.
\end{enumerate}

In traditional quantum measurement theory~\cite{Neumann}, the unitary process is considered as a process to entangle the measured system with the measuring apparatus before the projection process. The projection process gives definite final outcome. P2 essentially assumes the unitary process itself can have measurement result and can be assigned a (one-time) probability. \added{Zukowski and Markiewicz have} already pointed out that such assumption leads to a contradiction. But there is another problem with P2~\cite{Zukowski}. \added{The derivation in Ref. }\cite{Guerin} assumes that the joint probability $p(f_1,f_2)$ is derived through the standard probability axiom $p(f_1,f_2)=p(f_2|f_1)p(f_1)$, but does not give details on how the conditional probability is calculated in quantum theory. It is not clear how the unitary formulation presented in \cite{Guerin} can be applied to derive the conditional probability $p(f_2|f_1)$, because P2 assumes there is no ``collapse" after the first measurement. It is not a problem to compute the one-time probability $p(f_1)$ and $p(f_2)$. But in order to be able to calculate a two-time probability such as $p(f_2|f_1)$, one will have to apply the state update rule after the first measurement at time $t_1$, as shown in (\ref{twotimes}) for the two-time conditional probability. 

More crucially, even if we are able to calculate the conditional probability, there is still problem with P1 as P1 assumes the law of total probability, $p(f_1)=\sum_{f_2}p(f_1, f_2)$ is always true. We have shown in Theorem \ref{theorem1} and subsequent corollaries that the law of total probability is true in quantum theory only with certain conditions. The two POVM elements chosen in \cite{Guerin} are non-commutative, as shown in Eq.(17) in \cite{Guerin}. Thus, $p(f_1)=\sum_{f_2}p(f_1, f_2)$ does not necessarily hold. The proof in \cite{Guerin} assumes that $p(f_1)=\sum_{f_2}p(f_1, f_2)$ always holds based on P1, then deduces eventually that the two POVM elements should be commutative, and claims there is a contradiction. But such contradiction is due to the invalid assumption of $p(f_1)=\sum_{f_2}p(f_1, f_2)$ in P1, which in turn due to the fact that the two POVM elements are non-commutative. Since P1 is invalid, the contradiction does not lead to the desired conclusion that Wigner's friend cannot ``treat her perceived measurement outcome as having reality across multiple times".

\subsection{Relative facts, stable facts}
In relational interpretation of quantum mechanics (RQM) ~\cite{Rovelli96, Rovelli07, Transs2018, Rovelli18, Yang}, measurement result is considered meaningful only relative to the system that interacts with the measured system. A definite measurement result is referred as a fact. Quantum theory is about conditional probability for facts, given other facts. \added{Recently, Biagio and Rovelli} introduce the concept of \textit{stable fact} in the following sense~\cite{Rovelli21}. If, given the probability $p(a_i)$ for $N$ mutual exclusive facts $a_i (i=1\ldots N)$ and the conditional probability of another fact $b$, $p(b|a_i)$, the probability $p(b)$ (dropping index $j$ for $b_j$) is given by (\ref{totalLaw2}), then facts $a_i$ are considered stable. 

RQM states that fact is relative. Formally, if two systems $S$ and $F$ interact such that a variable $L_F$ of $F$ depends on the value of a variable $L_S$ of $S$, then the value of $L_S$ is said a fact relative to $F$~\cite{Rovelli21}. However, not all relative facts are stable. The main thesis of Ref.~\cite{Rovelli21} comprises two claims. First, The law of total probability (\ref{totalLaw2}) is satisfied only if $b$ and $a_i$ are facts relative to the \textit{same} system, say relative to system $F$; If $b$ is relative to another system $W\ne F$, (\ref{totalLaw2}) is not true in general. Mathematically, these can be expressed as
\begin{align}
    \label{stablefact1}
    p(b^{(F)}) &= \sum_i p(b^{(F)}|a_i^{(F)})p(a_i^{(F)}) \\
    \label{stablefact2}
    p(b^{(W)}) &\ne \sum_i p(b^{(W)}|a_i^{(F)})p(a_i^{(F)}).
\end{align}
Here it is important to label the reference system the fact is relative to. Second, if system $F$ goes through a decoherence process by interacting with an environmental system $E$, the resulting density matrix for $F$ is approximately given by $\rho=\sum_i\lambda_i|Fa_i\rangle\langle Fa_i|$, where $Fa_i$ is eigenvalues of $L_F$. Then, Eq.(\ref{stablefact2}) can be rewritten as 
\begin{equation}
    \label{stablefact3}
    p(b^{(W)}) = \sum_i p(b^{(W)}|Fa_i^{(E)})p(Fa_i^{(E)}).
\end{equation}
In such a case, facts $Fa_i$ relative to $E$ are stable for $W$.

Now let's examine the two claims more carefully. For the first claim, from Theorem \ref{theorem1}, Eq.(\ref{stablefact1}) is not necessarily true even if both $b$ and $a_i$ are facts relative to a same system.  That facts $b$ and $a_i$ are both relative to a same system means both facts are obtained through interactions with the same system, and the interactions can be represented by measurement operators $B$ and $A_i$, respectively. If $[B, A_i]=0$, Eq.(\ref{stablefact1}) is true. But there is no reason that $B$ and $A_i$ have to be commutative. If $[B, A_i]\ne 0$, Eq.(\ref{stablefact1}) is not true in general, unless other conditions such as condition $\textit{C2}^{\prime}$ or $\textit{C3}^{\prime}$ in Theorem \ref{theorem1} is satisfied. Indeed, the second claim (\ref{stablefact3}) is precisely the case where condition $\textit{C2}^{\prime}$ is met. Note that the reasoning from (\ref{stablefact1}) to (\ref{stablefact3}) is also applicable when facts $b$ and $a_i$ are relative to the same system but the corresponding measurement operators are non-commutative, $[B, A_i]\ne 0$.

Therefore, it is not clear that one can use the validity of the law of total probability (\ref{stablefact1}) and (\ref{stablefact3}) to distinguish stable facts from non-stable facts. Again, we are not opposite to the idea that facts are relative. What we are questioning here are the validity of (\ref{stablefact1}) without specifying the conditions, and the rigorousness of reasoning from (\ref{stablefact2}) to (\ref{stablefact3}). It appears that more careful investigation is needed in order to search for the criteria to define a ``stable" fact.

\section{Discussion and Conclusion}

\subsection{The Page-Wootters Timeless Formulation}
In the timeless formulation of quantum theory developed by Page and Wootters~\cite{PW}, time evolution is naturally emerged from quantum correlation between a clock and a system whose dynamics is tracked by the clock. \cite{Baumann} proposed several two-time formulations of conditional probability based on the Page-Wootters timeless mechanism. The advantage of such formulation is that from a timeless quantum state one can derive probability of a measurement event conditional on another event regardless of the temporary order of the two events.

Although the formulation in the present work is based on the regular time evolution dynamics in the Schrodinger picture, our definition of two-time conditional probability (\ref{twotimes}) is consistent with the definitions in ~\cite{Baumann}. For instance, for the case of two projection measurements $A_i$ and $B_j$ at $t_a$ and $t_b$, respectively, (\ref{twotimes}) gives the same transition probability (\ref{transProb}), as that in Eq. (29) of~\cite{Baumann}.

However, the timeless formulations of conditional probability in~\cite{Baumann} are applicable only to projection measurements, while the theory developed here is more general in the sense that it is applicable to POVM measurements. A two-time conditional probability formulation for projection measurements is insufficient to analyze the no-go theorems in \cite{Bong}. Moreover, our focus here is the validity of the law of total probability that is built on the definition of two-time conditional probability, which is missing in~\cite{Baumann} as its focus in only on the rules for two-time conditional probability.

It will be interesting to generalize the timeless Page-Wootters formulation of two-time conditional probability in~\cite{Baumann} to be able to handle POVM measurements. Although we expect such generalization should produce similar results as what are presented in this work.

\subsection{Limitations}
One limitation of the present works is that in Theorem \ref{theorem1}, we are only able to derive three sufficient conditions for the law of total probability to hold true. In theory, there can be many other sufficient conditions. It is desirable to find the \textit{sufficient and necessary} condition for the law of total probability to hold true in quantum theory. This remains a future investigation topic. Nevertheless, for the purpose of analyzing the EWF scenario and identifying the loopholes of the relevant no-go theorems, the conditions specified in Theorem \ref{theorem1} and subsequent Corollaries are sufficient.

\subsection{Conclusions}
In this paper, the standard rule to assign conditional probability in quantum theory, i.e., L\"{u}ders rule, is extended to include two-time POVM measurements. The extension is strictly based on the recursive application of the POVM measurement theory as shown in (\ref{POVM1}) and (\ref{POVM2}), and the assumption that probability distribution can be assigned only for completed quantum measurement. The resulting definition (\ref{twotimes}) is consistent with other works based on Page-Wootters formulation~\cite{Baumann} but with advantage of being able to apply to POVM measurements instead of just projection measurements.

More importantly, with the generalized two-time conditional probability formulation, we analyze the validity of the law of total probability. It is shown that the quantum version of the law of total probability does not hold true in general. Certain conditions related to the choice of measurement operators and the initial quantum state must be met in order for the law of total probability to hold. Specifically, such sufficient conditions are derived in Theorem \ref{theorem1} and Corollary \ref{corollary2}.

Application of the theory developed here to the Extended Wigner's Friend scenario reveals logical loopholes in several no-go theorems. These no-go theorems take for granted on the validity of the law of total probability (or, the law of marginal probability) in quantum theory. However, this is not the case as shown in Theorem \ref{theorem1} and Corollary \ref{corollary2}. Thus, the no-go theorems do not lead to the desired conclusions. For instance, the violation of the inequalities developed in \cite{Brukner} and \cite{Bong} in quantum theory does not necessarily lead to to desired statement that ``measured facts are observed-dependent". Instead, it just reconfirms the invalidity of the law of total probability or the law of marginal probability in quantum theory. We do not take a stand on the assertions themselves of the no-go theorems. It could be still a valid statement that ``measured facts are observed-dependent". What we show here is that there are logical loopholes to reach such statement. It is desirable to find more convincing proof and experimental testing because the implications of the extended Wigner's friend scenario are conceptually fundamental in quantum theory.



%
%

\added{
\begin{acknowledgements}
The author sincerely thanks the anonymous reviewers of this paper for their careful reviews. The valuable comments provided help to improve the clarity of the presentations and discussions.
\end{acknowledgements}
}


\begin{thebibliography}{}
\bibitem{Feynman}R. P. Feynman, Space-time approach to non-relativistic quantum mechanics. Rev. of Mod. Phys., 20, 367 (1948)
\bibitem{Zurek03}
Zurek, W. H.: Decoherence, Einselection, and the Quantum Origins of the Classical, Rev. of Mod. Phys. 75, 715 (2003)
\bibitem{Nielsen}Nielsen, M. A., and Chuang, I. L.: Quantum computation and quantum information. Cambridge University Press, Cambridge (2000)
\bibitem{Hayashi}M. Hayashi, S. Ishizaka, A. Kawachi, G. Kimura, and T. Ogawa, Introduction to Quantum Information Science, Springer, Heidelberg (2015)
\bibitem{Fine}A. Fine, Joint distributions, quantum correlations, and commuting observables. J. Math. Phys.
23, 1306–1310 (1982)
\bibitem{Malley}J. Malley and A. Fletcher, Joint distributions and quantum nonlocal Models. Axioms 3, 166-176 (2004)
\bibitem{Bobo}G. Bobo, Quantum conditional probability, PhD thesis, la Universidad Complutense de Madrid, 2010
\bibitem{Bobo2}G. Bobo, On Quantum Conditional Probability, An International Journal for Theory, History and Foundations of Science 28, 115 (2013)
\bibitem{Wigner}E. H. Wigner, Remarks on the mind-body question, in Symmetries and Reflections, pp 171-184 (Indiana University, 1967)
\bibitem{Wigner2}E. Wigner, The Scientist Speculates, edited by I. Good, pp. 284–302 (1961)
\bibitem{Deutsch}Deutsch, D.: Quantum theory as a universal physical theory. Int. J. Theo. Phys. 24, 1-41 (1985)
\bibitem{Brukner0}$\breve{\text{C}}$. Brukner, On the quantum measurement problem. In  Quantum [Un]speakables II; Bertlmann, R.,Zeilinger, A., Eds.; The Frontiers Collection; Springer: New York, NY, USA, (2016) 
\bibitem{Brukner}$\breve{\text{C}}$. Brukner, A no-go theorem for observer-independent facts, Entropy 20, 350 (2018)
\bibitem{Proietti19}M. Proietti, A. Picksron, F. Grattitti, P. Barrow, D. Kundys, C. Branciard, M. Ringbauer, and A. Fedrizzi, Experimental rejection of observer-independence in the quantum world, Science Advance 20 (5), 9, eaaw9832, arXiv:1902.05080 (2019).
\bibitem{Neumann}Von Neumann, J.: Mathematical Foundations of Quantum Mechanics, Chap. VI. Princeton University Press, Princeton Translated by Robert T. Beyer (1932/1955)
\bibitem{Bong}K-W Bong, A. Utreras-Alarcon, F. Ghafari, Y-C. Liang, N. Toschler, E. G. Cavalcanti, G. J. Pryde, and H. M. Wiseman, A strong no-go theorem on the Wigner's friend paradox. Nat. Phys., 16, 1199-1205 (2020)
\bibitem{Zukowski}M. Zukowski and M. Markiewicz, Physics and Metaphysics of Wigner’s Friends: Even Performed Pre-measurements Have No Results, Phys. Rev. Lett. 126, 130402 (2021)
\bibitem{Relano}A. Rela\~{n}o, Decoherent framework for Wigner's friend experiments. Phys. Rev. A 101, 032107 (2020)
\bibitem{Rovelli21}A. D. Biagio, C. Rovelli, Stable facts, relative facts. Found. of Phys. 51:30 (2021)
\bibitem{Guerin}P. A. Guerin, V. Baumann, F. DelSanto, and $\breve{\text{C}}$. Brukner, A no-go theorem for the persistent reality of Wigner's freind's perception. Nat. Comm. Phys. 4:93 (2021)
\bibitem{Luders}G. L\"{u}ders, ``Über die Zustandsanderung durch den Messprozess", Annalen der Physik 8: 322-328 (1951). English translation by Kirkpatrick, K. A.(2006):``Concerning the state-change due to the measurement process", Ann. Phys. (Leipzig) 15, No. 9 pp. 663-670. Also arXiv:0403007v2.
\bibitem{Cassinelli}G. Cassinelli, and P. Truni, ``Toward a Generalized Probability Theory: Conditional Probabilities" in G. Toraldo di Francia (ed.): Problems in the Foundations of Physics. Amsterdam- North Holland Publishing Company (1979)
\bibitem{Bub}J. Bub, ``Conditional Probabilities in Non-Boolean Possibility Structures" in C.A. Hooker (ed.) The Logic-Algebraic Approach to Quantum Mechanics,
Vol. II, pp. 209-226. The University of Western Ontario Series in Philosophy of Science, 5. Dordrecht, Holland: Reidel. (1979)
\bibitem{PW}D. N. Page and W. K. Wotters, Evolution without evolution: dynamics described by stational observables. Phys. Rev. D 27:2885 (1983)
\bibitem{Dolby}C. E. Dolby, The conditional probability interpretation of hamiltonian constraint, arxiv:0406034 (2004)
\bibitem{Lloyd}V. Giovannetti, S. Lloyd, and L. Maccone, Quantum time. Phys. Rev. D., 92 (4):045033 (2015)
\bibitem{Hohn}P. A. H\"{o}hn, Alexander R. H. Smith, and Maximilian P. E. Lock, Trinity of relational quantum dynamics, Phys. Rev. D 104, 066001 (2021)
\bibitem{Baumann}V. Baumann, F. D. Santo, A. R. H. Smith, F. Giacomini, E. Castro-Ruiz, and $\breve{\text{C}}$. Brukner, Generalized probability rules from a timeless formulation of Wigner's friend scenarios. Quantum 5, 524 (2021)
\bibitem{Busch}P. Busch, Quantum States and Generalized Observables: A Simple Proof of Gleason's Theorem. Phys. Rev. Lett. 91 (12): 120403 (2003)
\bibitem{Caves}C. M. Caves, C. A. Fuchs, K. K. Manne, and J. M. Renes, Gleason-Type Derivations of the Quantum Probability Rule for Generalized Measurements. Found. of Phys. 34, 193–209 (2004)
\bibitem{Hall}M. Hall, Relaxed Bell inequality and Kochen-Specker theorems. Phys. Rev. A 84, 022102 (2011)
\bibitem{Renner}D. Frauchiger, and R. Renner, Quantum theory cannot consistently describe the use of itself. Nat. Commun. 9, 3711 (2018)
\bibitem{Hall2}M. Hall, The significance of measurement independence for Bell inequalities and locality, In: At the Frontier of Spacetime, ed. T. Asselmeyer-Maluga, Springer, Switzerland, 189-204 (2016)
\bibitem{Rovelli96}
Rovelli, C.: Relational Quantum Mechanics, Int. J. Theor. Phys., 35, 1637-1678 (1996)
\bibitem{Rovelli07}
Smerlak M., and Rovelli, C.: Relational EPR, Found. Phys., 37, 427-445 (2007)
\bibitem{Transs2018}Transsinelli, M.: Relational Quantum Mechanics and Probability, Found. Phys., 48, 1092-1111 (2018)
\bibitem{Rovelli18}Rovelli, C.: ``Space is blue and birds fly through it", Phil. Trans. R. Soc. A 376, arXiv:2017.0312 (2018)
\bibitem{Yang}J. M. Yang, A Relational Formulation of Quantum Mechanics, Sci. Rep. 8:13305 (2018). J. M. Yang, Path integral implementation of relational quantum mechanics. Sci Rep 11, 8613 (2021)
\end{thebibliography}
\section*{Data Availability Statement}
The data that support the findings of this study are available within the article.


\appendix
\section{Proof of Theorem \ref{theorem1}}
With Condition $C1$, $[A_i, U^{\dag}B_jU] = 0$. Thus, $[\sqrt{A_i}, U^{\dag}B_jU] = 0$. The right hand side of (\ref{def3}) becomes $Tr(U^{\dag}B_jUA_i\rho)$. Given the completeness of POVM operators, $\sum_iA_i = I$, we have $\sum_i Tr(U^{\dag}B_jUA_i\rho) = Tr(U^{\dag}B_jU\rho)=Tr(B_j\rho(t_2))=p(b_j\text{ at }t_2|\rho)$. 

Given Condition $C2$, $\rho=\sum_k\lambda_k|\phi_k\rangle\langle \phi_k|$ and $A_i=|\phi_i\rangle\langle \phi_i|$, we get $\sqrt{A_i}\rho\sqrt{A_i}=\lambda_i|\phi_i\rangle\langle \phi_i|$. Then the right hand side of (\ref{def3}) becomes $Tr(U^{\dag}B_jU\lambda_i|\phi_i\rangle\langle \phi_i|)=\lambda_i\langle\phi_i|U^{\dag}B_jU|\phi_i\rangle$. The right hand side of (\ref{totalLaw5}) becomes $\sum_i\lambda_i\langle\phi_i|U^{\dag}B_jU|\phi_i\rangle$. On the other hand, $p(b_j\text{ at }t_2|\rho)=Tr(U^{\dag}B_jU\rho)=\sum_k Tr(U^{\dag}B_jU\lambda_k|\phi_k\rangle\langle \phi_k|)=\sum_i\lambda_i\langle\phi_i|U^{\dag}B_jU|\phi_i\rangle$, same as the right hand side of (\ref{totalLaw5}).

Given $\rho=|\Psi\rangle\langle\Psi|$, and $A_i$ is a projection operator, in Condition $C3$, we have $A_i^2=A_i$, and the right hand side of (\ref{totalLaw5}) becomes $\sum_i\langle\Psi|A_iU^{\dag}B_jUA_i|\Psi\rangle$. The left hand side $p(b_j\text{ at }t_2|\rho)=\langle\Psi|U^{\dag}B_jU|\Psi\rangle$. Again, by the completeness of POVM operators, $\sum_iA_i = I$, and we get $p(b_j\text{ at }t_2|\rho)=\sum_i\langle\Psi|U^{\dag}B_jUA_iA_i|\Psi\rangle$. To have both sides of (\ref{totalLaw5}) equal, we need
$\sum_i\langle\Psi|A_iU^{\dag}B_jUA_i|\Psi\rangle - \sum_i\langle\Psi|U^{\dag}B_jUA_iA_i|\Psi\rangle = 0$. This can be rearranged to $\sum_i\langle\Psi|[A_i, U^{\dag}B_jU]A_i|\Psi\rangle = 0$, and Condition $C3$ ensures this is the case.

\section{Proof of (\ref{C34})}
To avoid confusion, we need to restore the subscripts of operators as $C_i$ and $D_j$. Since $C_i$ and $D_j$ are projection operators, and $\rho=|\Psi\rangle\langle\Psi|$, the R.H.S. of (\ref{RHS7}) becomes $\sum_{ij}\langle\Psi|C_iD_jABC_iD_j|\Psi\rangle$. Given the completeness of $\{C_i\}$ and $\{D_j\}$, we have $\sum_{ij}C_iD_j=I$. Since $C_i^2=C_i$ and $D_j^2=D_j$, we further obtain $\sum_{ij}C^2_iD^2_j=I$. Then $p(ab  \text{ at }t_2|\rho, xy)=Tr(\rho AB)=\sum_{ij}\langle\Psi|ABC^2_iD^2_j|\Psi\rangle$. To make this equal to the R.H.S. of (\ref{RHS7}), one condition is to have $\langle\Psi|C_iD_jABC_iD_j|\Psi\rangle=\langle\Psi|ABC^2_iD^2_j|\Psi\rangle$. This is equivalent to $\langle\Psi|(C_iD_jABC_iD_j-ABC^2_iD^2_j)|\Psi\rangle=0$. But $(C_iD_jABC_iD_j-ABC^2_iD^2_j)=[C_iD_j, AB]C_iD_j$. Thus we have $\langle\Psi|[C_iD_j, AB]C_iD_j|\Psi\rangle=0$. Omitting the subscripts of $C_i$ and $D_j$ again gives (\ref{C34}).

\section{Non-commutation of Operators in ~\cite{Bong} }
The key characteristic of the EWF experiment is that the super-observer Alice (or Bob) performs measurements on the laboratory $L_1$ (or $L_2$) as a whole. Thus the measurement operator acts on both the observed system and the friend in the lab. \cite{Bong} carefully chooses the operators as following. When $x=1$, Alice's measurement is represented as $A(x=1)=|c\rangle\langle c|_{F_1}\otimes I_{s_1}$ where $c$ is the outcome Charlie obtains from his measurement on $s_1$ and $F_1$ refers to Charlie himself. For $x\in \{2,3\}$, Alice's measurement operator is $A(x)=U_{L_1}(I_{F_1}\otimes E^x_{s_1})U_{L_1}^{-1}$, where $U_{L_1}^{-1}$ is a unitary evolution that reverses the entanglement between $F_1$ and $s_1$, and $E^x_{s_1}$ is a positive operator on $s_1$ associated with outcome $a$ for measurement setting $x$. The operator associated with Charlie's measurement on $s_1$, according to ~\cite{Bong}, is described by an unitary operator $C(x) = U_{L_1}$ from Alice's perspective. $U_{L_1}$ acts on the same Hilbert space ${\cal{H}}_{F_1}\otimes {\cal{H}}_{s_1}$ and entangles $s_1$ and $F_1$. For $x=1$, $[A(x), C(x)]= |c\rangle\langle c|_{F_1}\otimes I_{s_1}U_{L_1} - U_{L_1}|c\rangle\langle c|_{F_1}\otimes I_{s_1}\ne 0$, and for $x\in \{2,3\}$, $[A(x), C(x)]=U_{L_1}(I_{F_1}\otimes E^x_{s_1}) - U^2_{L_1}(I_{F_1}\otimes E^x_{s_1})U_{L_1}^{-1} \ne 0$.

As already pointed out~\cite{Zukowski}, defining $C(x)$ as $U_{L_1}$ implies pre-measurement only with no measurement result, and leads to contradictions. An alternative choice of operation is that Charles performs a projection operation after the pre-measurement. This refines the definition of $C(x)$ to include both $U_{L_1}$ and a projection operation on $s_1$, i.e., $C(x)=U^{\dag}_{L_1}(I_{F_1}\otimes |c\rangle\langle c|)U_{L_1}$. With this refined definition, one can verify that $[A(x), C(x)]\ne 0$ still true. Choosing $C(x)=U^{\dag}_{L_1}(I_{F_1}\otimes |c\rangle\langle c|)U_{L_1}$ implies operator $C(x)$ is from Charlie's point of view. This may not be the original intention in \cite{Bong}. However, the key point here is that either choice of $C(x)$, $[A(x), C(x)]\ne 0$. The same analysis goes to operators $B(y)$ and $D(y)$, and the conclusion is that $[B(y), D(y)]\ne 0$ for $y\in \{1,2,3\}$.

\section{Proof That (\ref{totalLaw12}) Does Not Hold}
First we consider a simpler case that only Alice performs the two type of measurements and Bob does nothing. The law of total probability in this case can take the form of $p(a_2|\rho)=\sum_{a_1=\{0,1\}}p(a_2|a_1,\rho)p(a_1|\rho)$. We will show this is true due to the fact that the selected operators $A_1$ and $A_2$, and wave function (\ref{BruknerWF}) together meet condition $\textit{C3}^{\prime}$. To see this, substitute $A_1=|\phi_i\rangle\langle\phi_i|_{L_1}$ into condition $\textit{C3}^{\prime}$, $\textit{C3}^{\prime}$ becomes 
\begin{equation}
    \label{pureStateCondition3}
    \langle\varphi_i|[A_1,A_2]|\phi_i\rangle=0, \text{ where } |\varphi_i\rangle = \langle\Psi|\phi_i\rangle|\Psi\rangle,
\end{equation}
Now consider the case $a_1=0$, where $|\phi_0\rangle=|00\rangle_{L_1}$. From (\ref{BruknerWF}) one can calculate
\begin{align}
    \langle\phi_0|\Psi\rangle &= \frac{1}{\sqrt{2}}(-\sin\frac{\theta}{2}|00\rangle_{L_2}+\cos\frac{\theta}{2}|11\rangle_{L_2})\\
    |\varphi_0\rangle &= \langle\Psi|\phi_0\rangle|\Psi\rangle = \frac{1}{2}|00\rangle_{L_1}
\end{align}
Then from (\ref{nonCM}) and dropping the unimportant factor of $1/2$ for $|\varphi_0\rangle$, we have for the case of $a_1=0$
\begin{equation*}
    \begin{split}
    \langle\varphi_0|[A_1,A_2]|\phi_0\rangle &= \langle 00|(-1)^{a_2}(|00\rangle\langle 11| - |11\rangle\langle 00|)|00\rangle \\
    &=0.
    \end{split}
\end{equation*}
For the case of $a_2=1$, we can verify that $|\varphi_1\rangle=|11\rangle_{L_1}$ and
\begin{equation*}
    \begin{split}
    \langle\varphi_1|[A_1,A_2]|\phi_1\rangle &= \langle 11|(-1)^{1+a_2}(|00\rangle\langle 11| - |11\rangle\langle 11|)|00\rangle \\
    &=0.
    \end{split}
\end{equation*}
Therefore, condition $\textit{C3}^{\prime}$ is met with the choices of wavefunction and Alice's measurement operation. Similarly, if only Bob performs the two types of measurements and Alice does not perform any measurement, and let $B_1=|\phi'_i\rangle\langle\phi'_i|_{L_2}$, we can verify that
\begin{equation}
    \label{pureStateCondition4}
    \langle\varphi'_i|[B_1,B_2]|\phi'_i\rangle=0, \text{ where } |\varphi'_i\rangle = \langle\Psi|\phi'_i\rangle|\Psi\rangle.
\end{equation}
Thus, $p(b_2|\rho)=\sum_{b_1=\{0,1\}}p(b_2|b_1,\rho)p(b_1|\rho)$ holds true per Theorem \ref{theorem1}.

However, when we consider both Alice and Bob perform the measurements $\{A_1, A_2\}$ and $\{B_1, B_2\}$, respectively, the situation is different. By replacing operators $A, B, C, D$ in (\ref{C34}) with operators $A_2, B_2, A_1, B_1$, (\ref{C34}) reads
\begin{equation}
    \label{pureStateCondition5}
    \langle\Psi|[A_1B_1, A_2B_2]A_1B_1|\Psi\rangle=0.
\end{equation}
(\ref{pureStateCondition3}) and (\ref{pureStateCondition4}) together are not sufficient to ensure (\ref{pureStateCondition5}) is valid. Consequently, the law of total probability such as $p(a_2b_2|\rho)=\sum_{a_1,b_1=\{0,1\}}p(a_2b_2|a_1b_1,\rho)p(a_2b_2|\rho)$, is not valid. Let's confirm this by direct calculation for the case $a_2=0$ and $b_2=0$, where the corresponding projection operators are
\begin{align*}
    A_1 &=|\phi_i\rangle\langle\phi_i|_{L_1}, B_1 =|\phi_i\rangle\langle\phi_i|_{L_2}\\
    A_2&=\frac{1}{2}(|00\rangle_{L_1}+ |11\rangle_{L_1})(\langle 00|_{L_1}+ \langle 11|_{L_1}) \\
    B_2&=\frac{1}{2}(|00\rangle_{L_2}+ |11\rangle_{L_2})(\langle 00|_{L_2}+ \langle 11|_{L_2}).
\end{align*}
where $|\phi_0\rangle=|00\rangle$ and $|\phi_1\rangle=|11\rangle$. From (\ref{RHS7}), one can calculate that
\begin{align}
    &\sum_{a_1,b_1=\{0,1\}}p(a_2b_2|a_1b_1,\rho)p(a_2b_2|\rho)\\
    =&\sum_{a_1,b_1=\{0,1\}}\langle\Psi|A_1B_1A_2B_2A_1B_1|\Psi\rangle \\
    \label{C8}
    & = \sum_{i,j=\{0,1\}}|\langle\Psi|\phi_i\phi_j\rangle|^2 \langle\phi_i|A_2|\phi_i\rangle \langle\phi_j|B_2|\phi_j\rangle \\
    &= \frac{1}{4}.
\end{align}
Meanwhile, given $[A_2\otimes I_{L_2}, I_{L_1}\otimes B_2]=0$, the joint probability $p(a_2b_2|\rho)$ is well-defined as 
\begin{align*}
    &p(a_2=0,b_2=0|\rho)=Tr(A_2B_2\rho)=\langle\Psi|A_2B_2|\Psi\rangle\\
    &=\frac{1}{16}|\langle\Psi|(|00\rangle_{s_1C}+ |11\rangle_{s_1C})(|00\rangle_{s_2D}+ |11\rangle_{s_2D})|^2\\
    &=\frac{1}{8}\sin^2\frac{\theta}{2}.
\end{align*}
Thus, $p(a_2b_2|\rho)\ne\sum_{a_1,b_1=\{0,1\}}p(a_2b_2|a_1b_1,\rho)p(a_2b_2|\rho)$ for the case of $a_2=0$ and $b_2=0$. For other values of $a_2, b_2\in \{0, 1\}$, similar results can be calculated. Consequently, the law of marginal probability $p(a_2b_2|\rho)=\sum_{a_1,b_1=\{0,1\}}p(a_1b_1,a_2b_2|\rho)$ does not hold, if we define the joint probability $p(a_1b_1,a_2b_2|\rho)=p(a_2b_2|a_1b_1,\rho)p(a_2b_2|\rho)$.

If we add another condition that the quantum state $|\Psi\rangle$ is chosen as a product state of Hilbert space $\mathcal{H}_{L_1}$ and $\mathcal{H}_{L_2}$, then together with (\ref{pureStateCondition3}) and (\ref{pureStateCondition4}), (\ref{totalLaw12}) becomes true. To see this, let $|\Psi'\rangle=|\xi\rangle_{L_1}\otimes |\zeta\rangle_{L_2}$, (\ref{C8}) becomes
\begin{align}
    & \sum_{i,j=\{0,1\}}|\langle\xi|\phi_i\rangle\langle\zeta|\phi_j\rangle|^2 \langle\phi_i|A_2|\phi_i\rangle \langle\phi_j|B_2|\phi_j\rangle \\
    \label{C14}
    &=\sum_{i}|\langle\xi|\phi_i\rangle|^2\langle\phi_i|A_2|\phi_i\rangle\sum_j|\langle\zeta|\phi_j\rangle|^2\langle\phi_j|B_2|\phi_j\rangle.
\end{align}
But (\ref{pureStateCondition3}) implies $\sum_{i}|\langle\xi|\phi_i\rangle|^2\langle\phi_i|A_2|\phi_i\rangle=\langle\xi|A_2|\xi\rangle$, and (\ref{pureStateCondition4}) implies $\sum_j|\langle\zeta|\phi_j\rangle|^2\langle\phi_j|B_2|\phi_j\rangle=\langle\zeta|B_2|\zeta\rangle$. Thus, (\ref{C14}) becomes $\langle\xi|A_2|\xi\rangle\langle\zeta|B_2|\zeta\rangle = \langle\Psi'|A_2B_2|\Psi'\rangle=p(a_2b_2|\rho)$. This confirms (\ref{totalLaw12}) is valid.

However, $|\Psi\rangle$ in (\ref{BruknerWF}) is an entangled state between Hilbert space $\mathcal{H}_{L_1}$ and $\mathcal{H}_{L_2}$, so that (\ref{totalLaw12}) does not hold.  

\end{document}